\RequirePackage{fix-cm}
\documentclass[12pt]{article}
\usepackage{amsmath,amssymb,amsfonts}

\begin{document}
\title{Role of the constant deceleration parameter in cosmological models with perfect fluid and dark energy}

\author{D.D.Pawar , Y.S.Solanke* and R. V. Mapari**\\
School of Mathematical Sciences, Swami Ramanand Teerth\\
Marathwada University, Vishnupuri Nanded-431606 (India)\\
*Mungsaji Maharaj Mahavidyalaya,Darwha,Yavatmal 445202\\
** Dept. of Mathematics, Govt. Vidarbha Institute of\\ 
Science and Humanities,Amravati 444 604, India\\
E-mail: dypawar@yahoo.com, \\
yadaosolanke@gmail.com, r.v.mapari@gmail.com}

\maketitle
      Abstract: The main purpose of the present paper is to investigate LRS Bianchi type I metric in the presence of perfect fluid and dark energy. In order to obtain a deterministic solution of the field equations we have assumed that, the two sources of the perfect fluid and dark energy interact minimally with separate conservation of their energy momentum tensors. The EoS parameter of the perfect fluid is also assumed to be constant. In addition to these we have used a special law of variation of Hubble parameter proposed by Berman that yields constant deceleration parameter. For two different values of the constant deceleration parameters we have obtained two different cosmological models. The physical behaviors of both the models have been discussed by using some physical parameters.\\


   \section{Introduction}
   In the modern cosmology the numbers of recent astrophysical observational data suggest that the present universe is not only expanding but also accelerating and this accelerating phase of the universe is a recent phenomenon. Therefore naturally it is to be assumed that dark energy was insignificant in early evolution of the universe while it has the dominant contribution at the present accelerating epoch \cite{1}-\cite{4}. These observations also suggest that a transition of the universe from earlier deceleration phase to the accelerated stage of universe can be due to the domination of dark energy over other kinds of matter. In order to study the universe, the cosmologists have decided to obtain the large scale structure of the universe. In the formation of the large scale structure of the universe it should have decelerating expansion in early phase of matter era. Thus the formation of  structure in the universe is supported by decelerating model, also model should have decelerating as well as accelerating phase of universe to give a precise  form for this reasoning \cite{5}-\cite{7}. The simplest expanding cosmological models are those which are spatially homogeneous and isotropic. The evolution of isotropic cosmological models filled with perfect fluid dark energy have been extensively studied by many researchers. As it is predicted that the cosmic accelerated expansion of the universe is due to some kind of matter with negative pressure called the dark energy. The experimental observations such as cosmic microwave background radiation and large scale structure provide an indirect proof for the late time accelerated expansion of the universe \cite{8}-\cite{10}. The numbers of models have been studied by considering ordinary matter as a perfect fluid in the universe, but it is not sufficient to describe the dynamics of an accelerating phase of universe. This problem motivates the researchers to consider the models of the universe filled with dark energy along with perfect fluid \cite{11}-\cite{15}. In order to explain why the cosmic accelerated expansion of the universe happens, many candidates have been proposed. The cosmological constant is the prime candidate for dark energy even though having two well known problems such as the fine tuning and cosmic coincidence. The alternative candidates for the dark energy is  dynamical dark energy scenario. The quintessence, k-essence, chaplygin gas models, tachyon field, phantom field are some of the examples of dynamical dark energy models \cite{15a}-\cite{19}.Recently many authors studied magnetised dark energy models \cite{19a}-\cite{19f}.\\                                                    In the present paper we have studied LRS Bianchi type I model in the presence of perfect fluid and dark energy with variable EoS parameter for DE component, where as EoS parameter for perfect fluid is assumed to be constant. In order to obtain the exact solution of the Einstein’s field equations we have assumed a special law of variation of Hubble’s parameter that yields constant deceleration parameter. We have obtained two different cosmological models by using two different explicit forms of scale factors depending on the value of constant decelerating parameters. The paper is organized as follows. In section 2, the metric and the field equations have been presented. The exact solution of the field equations and physical behavior of the two different models is discussed in 3rd section. Finally we have concluded the opinion about the models in the 4th section.\\
   \section {Metric and the field equations }
              We consider LRS Bianchi type I metric \cite{19g}-\cite{20}  in the form given by

\begin{align}
ds^2=-dt^2+A^2(t)\{dx^2+dy^2+(1+\beta\int\frac{dt}{A^3})^2dz^2\}.
\end{align}
 Where $A(t)$  is a metric potential being a function of cosmic time $t$ and $\beta$  is positive constant. In natural units $(8\pi G=1,c=1)$ , the Einstein field equations in case of a mixture of perfect fluid and dark energy components as

 \begin{align}
  G_{ij}= R_{ij}-\frac{1}{2}Rg_{ij}=-T_{ij}
\end{align}
 Where $T_{ij}=T_{ij}^{(m)}+T_{ij}^{(de)}$  is the overall energy momentum tensor with $T_{ij}^{(m)}$   as the energy momentum tensors of the ordinary matter (perfect fluid) and $T_{ij}^{(de)}$   as the energy momentum tensor of the dark energy components which are respectively given by
\begin{equation}
T_j^{(m)i}=diag[-\rho^{(m)},p^{(m)},p^{(m)},p^{(m)}]=diag[-1,\omega^{(m)},\omega^{(m)},\omega^{(m)}]\rho^{(m)}
\end{equation}
and\\
\begin{equation}
 T_j^{(de)i}=diag[-\rho^{(de)},p^{(de)},p^{(de)},p^{(de)}]=diag[-1,\omega^{(de)},\omega^{(de)},\omega^{(de)}]\rho^{(de)}
\end{equation}
 Where ${\rho}^{(m)}$ and $p^{(m)}$  are the energy density and pressure of the perfect fluid components respectively whereas ${\rho}^{(de)}$  and $p^{(de)}$   are the corresponding energy density and pressure of the DE components, while  $\omega^{(m)}=\frac{p^{(m)}}{\rho^{(m)}}$    and  $\omega^{(de)}=\frac{p^{(de)}}{\rho^{(de)}}$   are the corresponding EoS parameters of the perfect fluid and dark energy.\\
 By assuming the commoving co-ordinate system,field equations (2) with the equations(3) and  (4) for the metric (1) turn into\\
\begin{align}
\frac{\dot{A}^2}{A^2}+2\frac{\ddot{A}}{A}=-\omega^{(m)}\rho^{(m)}-\omega^{(de)}\rho^{(de)},\\
3\frac{\dot{A}^2}{A^2}+\frac{2\beta\dot{A}}{A^4(1+\beta\int\frac{dt}{A^3})}=\rho^{(m)}+\rho^{(de)}.
\end{align}
By the equation of law of energy conservation (Bianchi identity)$ T_{;j}^{ij}=0$ we have\\
\begin{equation}
\dot{\rho}^{(m)}+3[1+\omega^{(m)}]H\rho^{(m)}+\dot{\rho}^{(de)}+3[1+\omega^{(de)}]H\rho^{(de)}=0
\end{equation}
\section{Solution of the Field Equations}\
The field equations (5) and (6) involve five unknown variables $A$,$\rho^{(m)}$,$\rho^{(de)}$,$\omega^{(m)}$,\\$\omega^{(de)}$.                                      Therefore in order to obtain the deterministic solution of the field equations we require three more suitable assumptions relating these unknown variables. According to Akarsu and Kilinc \cite{11}-\cite{13} , let us first assume that the perfect fluid and DE components interact minimally. Therefore equation of conservation of energy (7) can be split up into two separately additive conserved components which are as follow.\\
 The energy conservation equation of the perfect fluid $T_{;j}^{(m)ij}=0$   leads to
 \begin{equation}
 \dot{\rho}^{(m)}+3[1+\omega^{(m)}]H\rho^{(m)}=0.
 \end{equation}
 Similarly energy conservation equation of the DE components $T_{;j}^{(de)ij}=0$  leads to                                                                                                                \begin{equation}
 \dot{\rho}^{(de)}+3[1+\omega^{(de)}]H\rho^{(de)}=0.
 \end{equation}
 Here overhead dot represent the differentiation with respect to cosmic time where as superscripts (m) and (de) stand for perfect fluid (matter) and DE components respectively.\\                      Secondly we have assumed that the EoS parameter of the perfect fluid to be a constant.\\                           Thus \begin{equation}\omega^{(m)}=\frac{p^{(m)}}{\rho^{(m)}}=constant.\end{equation}                                                                                           Where as $\omega^{(de)}$  has been allowed to be a function of cosmic time.\\                                                                                                                                                              Since the line element (1) is completely characterized by Hubble parameter $H$, finally we have assumed that the mean generalized Hubble parameter $H$ is related to the scale factor $R$ by the relation given by \cite{21}.
 \begin{equation}
 H=lR^{-n}=l[A^3(1+\beta\int\frac{dt}{A^3})]^{\frac{-n}{3}}.
 \end{equation}
 Where $l>0$  and $n\geq0$ are constants.\\
Now we discuss the dark energy cosmological model for $n=0$ and $n\neq0$  by using equation (11) in following two respective subsections.
\subsection{DE Cosmological models for n=0}
Comparing equation (11) with the definition $H=\frac{\dot{R}}{R}$  and integrating we get
 \begin{equation}
 R(t)=k_1e^{lt},  l>0	
 \end{equation}
    Where $k_1$  is constant of integration.\\
 From given metric (1) the overall average scale factor $R$ is defined as	
 \begin{equation}
 R(t)=[A^3(1+\beta\int\frac{dt}{A^3})]^\frac{1}{3}.
 \end{equation}
 After little manipulations with the equations (12) and (13) we get
\begin{align}
A(t)=exp[lt+k_0e^{-3lt}],\\
(1+\beta\int\frac{dt}{A^3})=kexp[-3k_0e^{-3lt}].\\
Where\quad  k=k_1^3\quad  and\quad k_0=\frac{\beta}{9kl}
\end{align}
  Thus our required cosmological model for the given metric (1) takes the form                                                                                 \begin{equation}
ds^2=-dt^2+exp2[lt+k_0e^{-3lt}][dx^2+dy^2]+k^2exp[lt-2k_0e^{-3lt}]dz^2.
\end{equation}
\subsubsection{Some physical parameters of the model(17)}
 The directional Hubble parameters for the model along x , y and z axis are respectively given by
\begin{equation}
H_x=H_y=l(1-3k_0e^{-3lt}),\quad H_z=l(1+6k_0e^{-3lt}).
\end{equation}\\
Thus the mean generalized Hubble parameter for the model found to be
\begin{equation}
H=\frac{1}{3}(H_x+H_y+H_z)=l=constant.
\end{equation}\\
The mean anisotropy parameter $\triangle$ is defined and takes the value
\begin{equation}
\triangle=\frac{1}{3}\sum(\frac{H_i-H}{H})^2=18k_0^{2}e^{-6lt}.
\end{equation}
The spatial volume of the required model is obtained as
\begin{equation}
V=A^{3}(1+\beta\int\frac{dt}{A^3})=k^\frac{1}{3}e^{lt}.
\end{equation}
Similarly shear scalar as well as scalar expansion of the model are respectively given by
\begin{align}
\sigma^2=54k_0l^2e^{-6lt}\\
\quad and\quad  \theta=3H=3l=constant.
\end{align}
 The value of the constant deceleration parameter for this model is found to be
\begin{equation}
q=-\frac{R\ddot{R}}{\dot{R}^2}=-1
\end{equation}
The energy density of the perfect fluid by assuming its EoS parameter $\omega^{(m)}$  to be constant with the help of the equations (8), (13) and (14) is given by
\begin{equation}
\rho^{(m)}=k_2e^{-3[1+\omega^{(m)}]lt}.
\end{equation}
 Where $k_2$  being constant of integration.\\                                                                                                                                     The energy density of the DE component by using equation (6) with the equations (14), (15) and (25) is found to be                                                                                                                                                                             \begin{equation}
\rho^{(de)}=3l^2(1-9k_0^2e^{-6lt})-k_2e^{-3[1+\omega^{(m)}]lt}
\end{equation}
Similarly equation (5) by using the equations (14),(15),(25) and (26) EoS parameter for the DE component of model takes the value
\begin{equation}
\omega^{(de)}=-\frac{3l^2(1+9k_0^2e^{-6lt})+k_2\omega^{(m)}e^{-3[1+\omega^{(m)}]lt}}{3l^2(1-9k_0^2e^{-6lt})-k_2e^{-3[1+\omega^{(m)}]lt}}.
\end{equation}
\subsubsection{The physical behavior of the model (17)}
   The directional Hubble parameters $H_x=H_Y$  and $H_z$  are finite when the cosmic time is zero as well as infinity. Whereas the mean Hubble parameter is constant throughout the evolution of the universe. Therefore as rate of change mean Hubble parameter vanishes it indicates that the greatest value of mean Hubble parameter, the fastest rate expansion of the universe. Thus the model represent the inflationary era in the early universe and the very late time of the universe. The spatial volume of this model is finite when $t=0$   and expands exponentially as t increases and becomes infinitely large at $t=\infty$   . This shows that universe starts with constant volume and expands with exponential rate. The shear scalar $\sigma^2$ , the mean anisotropy parameter $\Delta$ are finite at$t=0$  and tends to zero as cosmic time tends to infinity. The energy density $\rho^{(m)}$  of the perfect fluid is constant $k_2$  when cosmic time is zero and decreases exponentially so as to converge at zero when EoS parameter of the perfect fluid $\omega^{(m)}\geq0$  as per the proposed assumption. But energy density $\rho^{(de)}$   of the DE component changes slightly when cosmic time is zero and decreases exponentially as time increases further it converges to nonzero constant $3l^2$  as well as DE component are finite at $t=0$  and tends to zero as cosmic time tends to infinity. In the present model ratio of  $\frac{\rho^{(de)}}{\rho^{(de)}+\rho^{(m)}}$  converges to 1 as t increases and this is sufficient to show that the dark energy dominates the perfect fluid in the inflationary era. The mean anisotropy parameter of expansion decreases monotonically when time increases and converges to zero when time is infinite. Also $\lim (\frac{\sigma^2}{\theta})=0$ when $ t\rightarrow  \infty$   indicates that model approach to isotropy for large value of cosmic time. The EoS parameter of the dark energy exhibits nontrivial behavior a the early time of the universe and converges to -1 for late time \cite{22}. Thus when cosmic time $t=\infty$  EoS parameter $\omega^{(de)}=-1$ . This is the simplest form of dark energy called vacuum energy, which is mathematically equivalent to the cosmological constant. But in some cosmological models value of the EoS parameter $\omega=-1$  is rejected so as to get the exact solution of the field equations \cite{23}.
\subsection{DE Cosmological Model for $n\neq0$}
Comparing equation (11) with the definition $H=\frac{\dot{R}}{R}$  and integrating we get                                                                                                                                                                  \begin{equation}
R(t)=(nlt+c_1)^\frac{1}{n},  l>0.
\end{equation}
Where $c_1$  is constant of integration.\\
From equation of a given metric (1) the overall average scale factor $R$  is defined as                                                                                                                                                                 \begin{equation}
R(T)=[A^3(1+\beta\int\frac{dT}{A^3})]^\frac{1}{3}
\end{equation}
Where, for the sake of simplicity we have chosen
\begin{equation}
T=nlt+c_1 \Rightarrow R(T)=T^\frac{1}{n}
\end{equation}
After little manipulation with the equations (28),(29) and (30) we get                                                                                                                                      \begin{align}
A(T)=T^\frac{1}{n}exp[\frac{\beta}{3(3-n)l}T^\frac{(n-3)}{n}],\\
[1+\beta\int\frac{dT}{A^3}]=exp[\frac{\beta}{(n-3)l}T^\frac{(n-3)}{n}].
\end{align}
 Thus our required cosmological model for the metric (1) becomes
\begin{equation}
ds^2=(\frac{-1}{n^2l^2})dT^2+[T^\frac{2}{n}exp\frac{2\beta}{3(3-n)l}T^\frac{n-3}{n}]\{dx^2+dy^2+[exp\frac{2\beta}{(n-3)l}T^\frac{(n-3)}{3}]dz^2\}.
\end{equation}
\subsubsection{Some physical parameters of the model (33)}
The directional Hubble parameters for the model(33) along x , y and z axis are respectively given  by
\begin{equation}
H_x=H_y=\frac{\dot{A}}{A}=\frac{1}{nT}-\frac{\beta}{3nlT^\frac{3}{n}}, H_z=\frac{\frac{d}{dT}[A(T)(1+\beta\int\frac{dT}{A^3})]}{A(T)(1+\beta\int\frac{dT}{A^3})}=\frac{1}{nT}+\frac{2\beta}{3nlT^\frac{3}{n}}.
\end{equation}
The mean generalized Hubble parameter for the model is

\begin{equation}
H=\frac{1}{3}(H_x+H_y+H_z)=\frac{1}{nT}.
\end{equation}
The mean anisotropy parameter $\triangle_1$  of the expansion for the model is                                                                \begin{equation}
\triangle_1=\frac{1}{3}\sum_{i=1}^3(\frac{H_{i}-H}{H})^2=\frac{2\beta^2}{9l^{2}T^\frac{2(3-n)}{n}}.
\end{equation}
The spatial volume $V$  of the model is found to be
\begin{equation}
V=A(1+\beta\int\frac{dT}{A^3})^\frac{1}{3}=T^\frac{1}{n}
\end{equation}
Similarly shear scalar $\sigma^2$  and scalar expansion $\theta$  are respectively found to be
\begin{equation}
\sigma^2=\frac{2\beta^2}{3n^{2}l^{2}T^\frac{6}{n}},  \theta=3H=\frac{3}{nT}.
\end{equation}
The value of the deceleration parameter for this model is obtained as
\begin{equation}
q=-\frac{R\ddot{R}}{\dot{R}^2}=n-1
\end{equation}
  As per our proposed assumption EoS parameter $\omega^{(m)}$  of perfect fluid being constant,energy density for the perfect fluid by using equation (8) with the equations (31) and (32) is given
\begin{equation}
\rho^{(m)}=c_2T^\frac{-3[1+\omega^{(m)}]}{n}.
\end{equation}
Where $c_2$ being constant of integration.\\
Equation (8) with the equations (31), (32) and (40) gives energy density of the DE component for the model (33) as
\begin{equation}
\rho^{(de)}=(\frac{1-2nl}{3n^2{l^2}})\frac{\beta^2}{T^\frac{6}{n}}+\frac{2(nl-1)\beta}{n^2{l}T^\frac{(n+3)}{n}}+\frac{3}{n^{2}T^{2}}-\frac{c_2}{T^\frac{3[1+\omega^{(m)}]}{n}}.
\end{equation}
Similarly equation (7) with the equations (31), (32), (40) and (41) gives EoS parameter of DE component as\\
\begin{equation}
\omega^{(de)}=-\frac{1}{\rho^{(de)}}\{\frac{(3-2n)}{n^{2}T^{2}}+\frac{\beta^2}{3n^{2}l^{2}T^\frac{6}{n}}+\frac{c_2\omega^{(m)}}{T^\frac{3[1+\omega^{(m)}]}{n}}\}.
\end{equation}
Where $\rho^{de}$ is given by the equation (41).
  \subsubsection{Physical behavior of the model (33)}
  The directional Hubble parameters $H_x=H_y,H_z$  and mean Hubble parameter $H$  are infinitely large at $T=0$ $(\because T=nlt+c_1)$     and becomes null when $T=\infty$ . It is observed that at $T=0$ , the spatial volume vanishes while all other parameters diverge. Thus the derived model starts evolving with zero volume and expands with cosmic time. This singularity is point type because metric potential $A(T)$  vanishes at the initial moment. The mean anisotropy parameter $\triangle_1$ , the expansion scalar $\theta$  and shear scalar $\sigma^2$  all vanish when $T\rightarrow\infty$ , which indicates that universe is expanding with increase in cosmic time. Also $\lim\frac{\sigma^2}{\theta}=0$ when $T\rightarrow\infty$, provided $n<6$  which shows that the model approaches isotropic for large value of cosmic time. According to Collins and Hawking (1973) \cite{24} all the candidates for homogeneity and isotropizations are satisfied by the present model. Thus the present model approaches isotropic during the late time of its evolution.
  \section{Conclusion}
  In t he present paper in order to obtain the exact solution of the Einstein field equations of the LRS Bianchi type I model we have assumed the special law of variation of Hubble’s parameter that yields the constant deceleration parameter. As we have defined the Hubble’s parameter in equation (11) gives rise to two types of cosmological models depending on the nature of the value of constant deceleration parameter whether it is positive or negative. The first form of the universe having negative value of deceleration parameter shows the exponential expansion of the universe while second form of the universe having positive value of deceleration parameter shows the power law expansion of the universe. For the exponential expansion model all the parameters $H_x$,$H_y$,$H_z$,$\triangle$,$\sigma^2$  are constant at $t=0$  . The mean Hubble parameter as well as expansion scalar of the universe is uniform throughout the evolution of the universe. As $t\rightarrow\infty$  the EoS parameter of the DE component is $-1$  i.e. $\omega^{(de)}=-1$ which may be considered as vacuum energy density. Obviously it is equivalent to cosmological constant and it is important to note that this class of solution is consistent with the recent observations of the supernova Ia  [1-4]. For the power law expansion model $H_x$,$H_y$,$H_z$,$\triangle_1$,$\sigma^2$,$\theta$  all these parameters are infinitely very large at initial moment and decreases with increase in time and vanishes at large value of cosmic time. Thus from the derived two cosmological models we lead to the conclusion that if $q<0$  the model expand exponentially and later accelerates where as when $q>0$ model expands but decelerates.


\begin{thebibliography}{References}
\bibitem{1}
Perlmutter, S. etal.  : Nature 391, 51 (1998).
\bibitem{2}
 Reiss,A.etal. : Astrophys. J. 517, 565 (1999).
 \bibitem{3}
Knop, R. A. etal. : Astrophys. J. 598, 102 (2003).
 \bibitem{4}
Reiss, A. etal. : Astron. J. 607, 665 (2004).                                                                                                                                                               \bibitem{5}
Caldwell, R. R. : Phys. Lett. B 545, 23 (2002).
\bibitem{6}
Caldwell, R. R., Doran, M. : Phys. Rev. D 69, 103517 (2004).
\bibitem{7}
Haung, Z. Y. , Wang, B., Abdulla, E. Sul, R. K. : J. Cosmol. Astropart. Phys. 05, 013
      (2006).
\bibitem{8}
 B. Jain and A. Taylor, : Phys. Rev. Lett. 91, 141302 (2003).
\bibitem{9}
P. Peebles and B. Ratra, : Rev. Mod. Phys. 75, 559 (2003).
\bibitem{10}
M. Tegmark, etal. : Phys. Rev. D69, 103501 (2004).
\bibitem{11}
 Akarsu.Ozgur,Kilink,C.B.:Gen. Relativ. Gravit, 42, 119 (2010a).                                                                                                           \bibitem{12}
 Akarsu.Ozgur, Kilink,C.B.: Gen.Relativ. Gravit, 42, 763 (2010b).
 \bibitem{13}
 Akarsu.Ozgur, Kilink,C.B.: Astrophys.Space Sci.326, 315(2010c).
\bibitem{14}
 Yadav, A. K. Yadav, V. L. : Int. J. Theor. Phys. 50, 218 (2011).
\bibitem{15}
 Pawar, D. D., Pawar, S. S.:Prespacetime Journal, August 2015, Volume 6 , Issue 8 , pp. 719-732.
\bibitem{15a}
 Ratra, B.,Peebles, J. : Phys. Rev.D 37, 321 (1988).
\bibitem{16}
Chiba,T.,Okabe,T.,Yamaguchi,M.:Phys. Rev. D62,023511(2000).
\bibitem{17}
 Kamenshchik, A.,Moschella,U.,Pasquier, V. : Phys. Lett. B511, 265 (2001).
\bibitem{18}
Padmanabhan,T. : Phys. Rev. D66, 021301 (2002).
\bibitem{19}
	Pawar,D.D.,Solanke,Y.S.:International Journal of Scientific and Innovative Mathematical Research (IJSIMR)
Volume 3, Special Issue 3, July 2015, PP 789-794.
\bibitem{19a}
S. D. Katore, M. M. Sancheti, and S. P. Hatkar: Int. J. Mod. Phys. D 23, 1450065 (2014).
\bibitem{19b}
S. D. Katore, A. Y. Sheikh: The African Review of Physics (2014) 9:0035, 269-276.
\bibitem{19c}
D.D.Pawar,Y.S.Solanke,:Advances in High Energy Physics,Volume 2014,Article ID 859638, 9 pages http://dx.doi.org/10.1155/2014/859638
\bibitem{19d}
D.D.Pawar,Y.S.Solanke,S. P. Shahare:Bulg. J. Phys. 41 (2014) 60–69.
\bibitem{19e}
V. R. Chirde, S. H. Shekh:International Journal of Advanced Research (2014), Volume 2, Issue 6, 1103-1114
\bibitem{19f}
Kanika Das,Tazmin Sultana:Astrophys Space Sci (2015) 357:118.
\bibitem{19g}
Abdussattar and S.R.Prajapati, : Astrophys.Space Sci.331,657 (2011).
\bibitem{20}
Pawar, D. D., Dagwal, V.J.:Int.J.Theor.Phys. DOI ;10.1007/s10773-014-2043-7.
\bibitem{21}
Berman,M.S. : Nuovo Cimento B.74, 182 (1983).
\bibitem{22}
Pawar,D.D.,Solanke,Y.S.:Int.J.Theor.Phys. DOI;10.1007/s10773-014-2101-1.
\bibitem{23}
Pawar,D.D.,Solanke,Y.S.,Bayskar, S.N.:Prespacetimeary2014/Volume5/Issue2/pp.60-68.
\bibitem{24}
Collins, C.B., Hauking, S.W.. : Astrophys. J. 180, 317 (1973).
\end{thebibliography}
\end{document}